\documentclass[aps,prb,twocolumn,nobibnotes,showpacs,amsmath,amssymb,superscriptaddress]{revtex4-1}
\usepackage{graphicx}% Include figure files
\usepackage{dcolumn}% Align table columns on decimal point
\usepackage{epstopdf}
\usepackage{bm}% bold math
\usepackage{color}

\bibpunct{[}{]}{,}{n}{}{}

\begin{document}
\title{Electronic properties of single-layer antimony: Tight-binding model, spin-orbit coupling and the strength of effective Coulomb interactions}

\author{A.~N. Rudenko}
\email[]{a.rudenko@science.ru.nl}
\affiliation{\mbox{Institute for Molecules and Materials, Radboud University, Heijendaalseweg 135, NL-6525 AJ Nijmegen, The Netherlands}}
\author{M.~I. Katsnelson}
\affiliation{\mbox{Institute for Molecules and Materials, Radboud University, Heijendaalseweg 135, NL-6525 AJ Nijmegen, The Netherlands}}
\author{R. Rold\'{a}n}
\affiliation{Instituto de Ciencia de Materiales de Madrid, ICMM-CSIC, Cantoblanco, E-28049 Madrid, Spain}
\date{\today}

\begin{abstract}
The electronic properties of single-layer antimony are studied by a combination of first-principles and tight-binding methods. The band structure obtained from relativistic density functional theory is used to derive an analytic tight-binding model that offers an efficient and accurate description of single-particle electronic states in a wide spectral region up to the mid-UV. The strong ($\lambda=0.34$ eV) intra-atomic spin-orbit interaction plays a fundamental role in the band structure, leading to splitting of the valence band edge and to a significant reduction of the effective mass of the hole carriers.  To obtain an effective many-body model of two-dimensional Sb we calculate the screened Coulomb interaction and provide numerical values for the on-site $\bar{V}_{00}$ (Hubbard) and intersite $\bar{V}_{ij}$ interactions. We find that  the screening effects originate predominantly from the 5$p$ states, and are thus fully 
captured within the proposed tight-binding model.
The leading kinetic and Coulomb energies are shown to be comparable in magnitude, $|t_{01}|/(\bar{V}_{00}-\bar{V}_{01})$~$\sim$~1.6, which
suggests a strongly correlated character of 5$p$ electrons in Sb.
The results presented here provide an essential step toward the 
understanding and rational description of a variety of electronic properties of this two-dimensional material.
\end{abstract}

\maketitle

Single layers of antimony crystal (SL-Sb) have been recently produced
using different methods, including mechanical exfoliation \cite{Ares},
liquid-phase exfoliation \cite{Gibaja,Huo}, and epitaxial growth on a substrate \cite{JJi,Kim,Lei}.
Two-dimensional (2D) antimony complements the list of elemental 2D materials 
available to experiment, among which are graphene \cite{Novoselov} with its 
group IV analogs silicene \cite{Vogt} and germanene \cite{Zhang}, 
few-layer black phosphorus \cite{Li}, as well as the more exotic materials
stanene \cite{Zhu} and borophene \cite{Mannix}. The presence of a moderate 
band gap and excellent 
environmental stability \cite{Ares} combined with predictions of 
a reasonable carrier mobility \cite{Pizzi} makes 2D antimony a 
promising candidate for electronic, transport, and optical applications, as
well as for the realization of topological phase transitions \cite{Kim}.

Theoretically, electronic properties of SL-Sb have been  
studied using first-principles methods \cite{Pizzi,SZhang,Wang,Akturk,Xu,Singh}. 
In many cases, however, the performance of such methods turns out to be 
limited by high computational cost, which prevents one from reliably describing 
the properties of realistic materials, especially at large scales and beyond
single-particle approximations. The method of model Hamiltonians is an alternative
approach to address the problem of the electronic structure, which is less
transferable, but more efficient and flexible. Among 2D materials, several tight-binding (TB)
models have been proved to capture the relevant electronic states in
 graphene \cite{Reich,Gruneis} and 
its derivatives \cite{Mazurenko},
transition metal dichalcogenides \cite{Rostami,Cappelluti,Zahid,Xiao} and different phases of 
phosphorus \cite{Takao, Rudenko1,Rudenko2, Mogulkoc},
while single-layer antimony is still missing from the list. 

Another important ingredient for a reliable physical description of materials is
the information on 
the strength of
the Coulomb interaction, which
directly affects optical properties and plays a key role in phenomena such as charge carrier scattering 
and superconductivity. 
Besides, the Coulomb interaction is an important component of many-body theory,
which is aiming at providing an exact solution to the electronic structure problem. 
To date, the problem of the Coulomb interactions and their screening beyond the long-wavelength limit 
has only scarcely been addressed in the context of 2D materials \cite{Mazurenko,Wehling,Schuler}.

In this Rapid Communication, we derive a tractable TB 
model for SL-Sb, which can serve as a starting point for a 
comprehensive analysis of electronic properties including many-body
effects as well as for large-scale simulations of this material.
We explicitly take into account spin-orbit (SO) coupling, whose effect in the band structure is discussed,
and estimate the strength of the Coulomb interaction in SL-Sb. Our results suggest a strongly correlated character of 5$p$ electrons in SL-Sb. We show that the proposed analytical TB model captures the dominant contribution of the 
screening effects and thus can be considered as a complete description of 
the electronic states in SL-Sb in the spectral region up to the mid-UV.

Equilibrium structural parameters and reference electronic bands have 
been obtained at the density functional (DFT) level using the {\sc vasp} code \cite{Kresse1,Kresse2}. 
The generalized gradient approximation \cite{PBE} was used in combination with 
the projected augmented-wave method \cite{PAW}. The kinetic energy cutoff was
set to 200 eV, the vertical interlayer separation to 30 \AA, and the Brillouin zone 
sampled by a ($48\times48$) {\bf k}-point mesh. An energy window of $\sim$50 eV
was used in the polarizability calculations. All the results are
checked for numerical convergence. The construction of the Wannier functions 
and TB parametrization of the DFT Hamiltonian are done with
the {\sc wannier90} code \cite{wannier90}.

\begin{figure}[tbp]
\includegraphics[width=0.47\textwidth, angle=0]{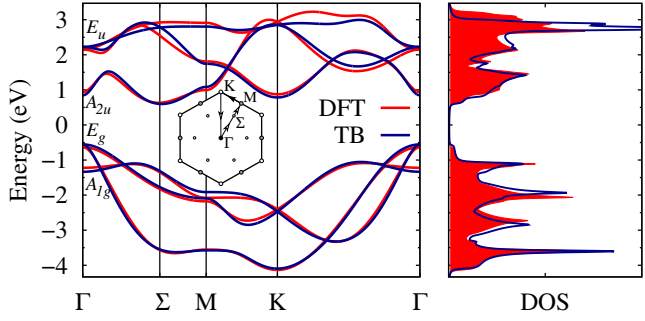}
\caption{Band structure (left) and density of states (right) calculated
without SO coupling for SL-Sb using the DFT and TB model [Eq.~(\ref{tb_hamilt})] presented 
in this work. Irreducible representations of the $D_{3d}$ point group \cite{Kogan2} realized for the states at the $\Gamma$ point are indicated.}
\label{bands}
\end{figure}

Single-layer Sb adopts a buckled honeycomb structure (space group $D^3_{3d}$)
with the lattice parameter $a=4.12$ \AA~and two sublattices vertically 
displaced by $b=1.65$ \AA.
Structurally, Sb layers are similar to silicene \cite{Silicene} or germanene \cite{Germanene} yet with a 
larger buckling $b$, which is comparable to that predicted for single layers of
the $A7$ (blue) phase of elemental phosphorus \cite{Tomanek}.
If SO coupling is neglected, the top of the valence band located at the $\Gamma$ point is doubly degenerate (Fig.~\ref{bands}) for each spin channel. 
The corresponding states ($E_g$) are composed of the $p_x$, $p_y$ orbitals only and are symmetric with respect to the inversion center. 
In contrast, the bottom of the conduction band is shifted from the $\Gamma$ to a 
$\Sigma$ point in the $\Gamma$-$M$ direction by $\sim$2/3 of the $\Gamma$-$M$ distance. Orbital decomposition of the corresponding wavefunction at $\Sigma$ yields 
$|\psi^{\mathrm{CB}}(\Sigma)\rangle \approx 0.14|s \rangle + 0.61|p_z\rangle + 0.76 |p_{x,y}\rangle$.
An indirect gap between the $\Gamma$ and $\Sigma$ points is estimated to be $\sim$1.26 eV. 

\begin{figure}[tbp]
\includegraphics[width=0.50\textwidth, angle=0]{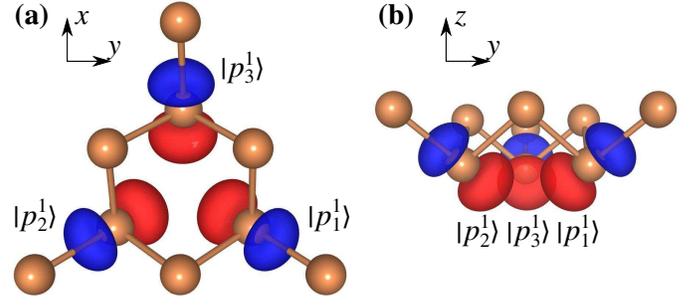}
\caption{Wannier orbitals of SL-Sb corresponding to the basis of the TB Hamiltonian presented in this work.
For clarity, orbitals are shown for one sublattice ($k=1$) with one orbital per atom only. The orbitals in the second
sublattice are symmetric with respect to the inversion center.}
\label{orbitals}
\end{figure}

The fact that the hole and electron states are symmetry inequivalent makes the construction of a simple low-energy TB model for Sb not trivial. 
However, given that the valence and conduction bands have a predominantly $p$ character, and that they are separated from other states,
it turns out to be possible to provide an accurate description of those states in terms of a tractable TB model in the whole energy region.
The parametrization procedure used in our work is based on the formalism
of maximally localized Wannier functions (WFs) \cite{Marzari,MLWF,Lado}. 
In this formalism, the cell periodic part $u^{H}_{n\bf k}({\bf r})$ of the 
Bloch functions 
$\psi^{H}_{n\bf k}({\bf r})=u^{H}_{n\bf k}({\bf r})e^{{i{\bf k}\cdot}{\bf r}}$,
representing the eigenfunctions of the first-principles Hamiltonian 
$H^{H}(\bf k)$, transforms according to
\begin{equation}
u^{W}_{n\bf k}({\bf r}) = \sum_{m} U^{\bf k}_{mn} u^{H}_{m\bf k}({\bf r}),
\label{gauge}
\end{equation}
where $n$ is the band index and ${\bf k}$ is the crystal momentum. In 
Eq.~(\ref{gauge}), $U_{mn}^{\bf k}$ is a unitary matrix chosen so that it 
minimizes the spread of the Wannier orbitals
$
w_{n{\bf R}_i}({\bf r}) = \frac{1}{N_k} \sum_{\bf k} e^{-i{\bf k}\cdot{\bf R}_i} \psi^{W}_{n{\bf k}}({\bf r})
$
centered at ${\bf R}_i$ \cite{footnote2}.
In the case of SL-Sb, the relevant bands (Fig.~\ref{bands}) are 
isolated, thus the construction of WF does not require a disentanglement 
procedure, which makes the WFs uniquely defined within the scheme
of maximal localization. A real-space distribution of the WFs obtained for
SL-Sb is shown in Fig.~\ref{orbitals}. They represent a combination
of three $p$-like orbitals localized on each Sb atom, giving rise
to six WFs per cell. In terms of the atomiclike orbitals $|p_x \rangle$, $|p_y \rangle$, and $|p_z \rangle$,
the corresponding WFs can be expressed for each atomic site as
\begin{eqnarray}
\notag
|p^{(k)}_1 \rangle = \mathrm{sin}\alpha \left[(-1)^{k+1}\frac{1}{2} |p_x \rangle + \frac{\sqrt 3}{2}|p_y \rangle \right] + (-1)^k \mathrm{cos}\alpha \, |p_z \rangle, \\
\notag
|p^{(k)}_2 \rangle = \mathrm{sin}\alpha \left[(-1)^{k+1}\frac{1}{2} |p_x \rangle - \frac{\sqrt 3}{2}|p_y \rangle \right] + (-1)^k \mathrm{cos}\alpha \, |p_z \rangle, \\
|p^{(k)}_3 \rangle = \mathrm{sin}\alpha \, (-1)^{k} |p_x \rangle + (-1)^k \mathrm{cos}\alpha \, |p_z \rangle ,\quad \quad \quad \quad \quad \quad \quad \, \, \,
\label{Tmatrix}
\end{eqnarray}
where $k$ is the sublattice index (1 or 2), and 
$\alpha=\mathrm{arccos}(1/{\sqrt{1 + a^2 / 3 b^2}})\approx55.3^{\mathrm{o}}$ 
is the angle formed by the inclination of an orbital from the $z$ direction. All three orbitals
are equivalent and symmetry related.

The resulting nonrelativistic TB model is given by an effective Hamiltonian,
\begin{equation}
H_{0}=\sum_{mn}\sum_{ij} t^{mn}_{ij} c_{im}^{\dag}c_{jn},
\label{tb_hamilt}
\end{equation}
where $t^{mn}_{ij}$ is the effective hopping 
parameter describing the interaction between $m$ and $n$ orbitals
residing at atoms $i$ and $j$, respectively. In Eq.~(\ref{tb_hamilt}),
$c_{im}^{\dag}$ ($c_{jn}$) is the creation (annihilation) operator of 
electrons at atom $i$ ($j$) and orbital $m$ ($n$). To make the model
more tractable yet accurate enough, we ignore long-range hopping parameters with amplitudes
$|t|$~$<$~25 meV.
\begin{figure}[tbp]
\includegraphics[width=0.45\textwidth, angle=0]{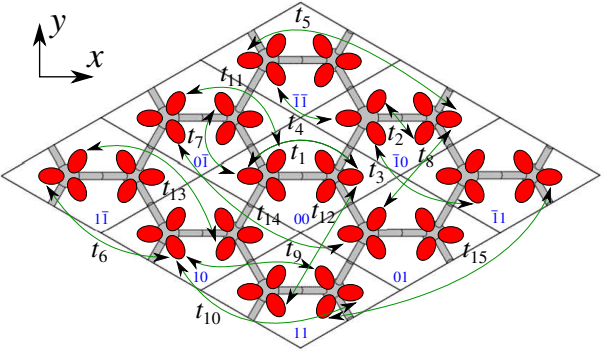}
\caption{Schematic representation of the crystal structure (top view) and 
relevant hopping parameters ($t_i$)
involved in the TB model of SL-Sb. Interacting orbitals are 
depicted by red ovals, corresponding to the positive part of the Wannier orbitals (cf. Fig.~\ref{orbitals}).
The hopping amplitudes are given in \mbox{Table \ref{hoppings_table}}. Blue labels mark relative unit cell coordinates.}
\end{figure}
    \begin{table}[b]
    \centering
    \caption[Bset]{Hopping amplitudes $t_i$ (in eV) assigned to the TB Hamiltonian [Eq.~(\ref{tb_hamilt})] of SL-Sb. $d$ denotes the 
                   distance between the lattice sites on which the interacting orbitals are centered.
                   $N_c$ is the corresponding coordination number. The hoppings are schematically shown in Fig.~\ref{hoppings}.}
\label{hoppings}
 \begin{tabular}{cccccccccccccc}
      \hline
      \hline
\cline{2-4}
\cline{6-8}
\cline{11-13}
 $i$&  $t_i$(eV) &  $d$(\AA) &  $N_c$  & $i$&  $t_i$(eV) &  $d$(\AA) &  $N_c$ &     & $i$ &  $t_i$(eV) &  $d$(\AA)  &  $N_c$ \\
     \hline
   1&    -2.09    &    2.89    &    1    &  6 &     0.21    &    4.12    &    1   &     &  11 &   -0.06    &    4.12    &   2  \\
   2&     0.47    &    2.89    &    2    &  7 &     0.08    &    2.89    &    2   &     &  12 &   -0.06    &    5.03    &   1  \\
   3&     0.18    &    4.12    &    4    &  8 &    -0.07    &    5.03    &    2   &     &  13 &   -0.03    &    6.50    &   2  \\
   4&    -0.50    &    4.12    &    1    &  9 &     0.07    &    6.50    &    2   &     &  14 &   -0.04    &    8.24    &   1  \\
   5&    -0.11    &    6.50    &    2    &  10&     0.07    &    6.50    &    2   &     &  15 &   -0.03    &    8.24    &   1  \\
      \hline                  
      \hline
    \end{tabular}
\label{hoppings_table}
    \end{table}
The orbitals and the
relevant hopping
parameters are schematically shown in Fig.~\ref{hoppings}. 
In reciprocal space, the Hamiltonian matrix can be represented as
\begin{equation}
H({\bf k})=
\begin{pmatrix}
E({\bf k})         &  T({\bf k}) \\
T^{\dag}({\bf k})  &  E({\bf k}_r)
\end{pmatrix},
\label{H}
\end{equation}
where $E({\bf k})$ and $T({\bf k})$ are $3\times3$ matrices describing the
intrasublattice and intersublattice interactions, respectively. The
corresponding matrices have the form,
\begin{equation}
E({\bf k})=
\begin{pmatrix}
A(\bar{\bf k})         &  B({\bf k})           & B^{*}(\bar{\bar{\bf k}}) \\
B^{*}({\bf k})    &  A(\bar{\bar{\bf k}})         & B(\bar{\bf k}) \\
B(\bar{\bar{\bf k}})  &  B^{*}(\bar{\bf k})  & A({\bf k})
\end{pmatrix},
\label{E}
\end{equation}
and
\begin{equation}
T({\bf k})=
\begin{pmatrix}
C({\bf k})      &    D(\bar{\bf k})  &   C(\bar{\bar{\bf k}}) \\
D(\bar{\bar{\bf k}})    &    C({\bf k})    &   C(\bar{\bf k}) \\
C(\bar{\bf k})    &    C(\bar{\bar{\bf k}})  &   D({\bf k})
\end{pmatrix}.
\label{T}
\end{equation}
In Eqs.~(\ref{E}) and (\ref{T}), $\bar{\bf k}$ ($\bar{\bar{\bf k}}$) is the ${\bf k}$ vector rotated
by $2\pi/3$ ($4\pi/3$), whereas the subscript $r$ 
of ${\bf k}$ in Eq.~(\ref{H}) indicates rotation in the opposite direction, equivalent to the vertical mirror symmetry operation ($\sigma_d$).
The matrix elements appearing in
Eqs.~(\ref{E}) and (\ref{T}) read
\begin{equation}
A({\bf k}) = 4t_3\,\mathrm{cos}\left(\frac{{\sqrt 3}}{2}k_xa\right)\mathrm{cos}\left(\frac{1}{2}k_ya\right) + 2t_{11}\,\mathrm{cos}\left(k_ya\right) ,
\end{equation}
\begin{equation}
B({\bf k}) = t_4\,e^{ik_ya} + t_6\,e^{-ik_ya} + t_{14}\,e^{2ik_ya} + t_{15}\,e^{-2ik_ya} ,
\end{equation}
\begin{multline}
C({\bf k})=2t_7\,e^{i\frac{\sqrt 3}{6}k_xa}\mathrm{cos}\left( \frac{1}{2}k_ya \right) + 2t_{8}\,e^{-i{\frac{{\sqrt 3}}{3}}k_xa}\mathrm{cos}(k_ya) + \quad \quad \quad \\
2t_{10}\,e^{i\frac{{\sqrt 3}}{6}k_xa}\mathrm{cos}\left( \frac{3}{2}k_ya \right) + t_{12}\,e^{i\frac{2\sqrt 3}{3}k_xa} , 
\end{multline}
\begin{multline}
D({\bf k}) = t_1\,e^{-i\frac{\sqrt 3}{3}k_xa} + 2t_2\,e^{i\frac{{\sqrt 3}}{6}k_xa}\mathrm{cos}\left(\frac{1}{2}k_ya\right) + \\ 
2t_5\,e^{-i\frac{5{\sqrt 3}}{6}k_xa}\mathrm{cos}\left( \frac{1}{2}k_ya  \right) + 2t_9\,e^{i\frac{2\sqrt 3}{3}k_xa}\mathrm{cos}(k_ya) + \\
2t_{13}\,e^{i\frac{\sqrt 3}{6}k_xa}\mathrm{cos}\left(\frac{3}{2}k_ya\right)  .
\end{multline}
The resulting band structure and density of states (DOS) calculated with the given TB model is shown in Fig.~\ref{bands}, from which one can see a very good match between the TB and
original first-principles spectra. The agreement in the low-energy region can be quantified by the effective masses, which are accurately reproduced by the TB model as
shown in Table \ref{masses}. Interestingly, for all relevant effective masses $m<1/{\sqrt 3}m_0$ holds, which according to the Landau-Peierls theory suggests that in the presence of
a perpendicular magnetic field at low temperatures, charge carriers in SL-Sb would respond diamagnetically, contrary to graphene \cite{Stauber}.
We note, however, that many-body effects not considered here might renormalize the dispersion $E_{\bf k} = \varepsilon_{\bf k} + \mathrm{Re}\Sigma({\bf k},E_{{\bf k}})$, and enhance the quasiparticle effective masses \cite{DasSarma}.

    \begin{table}[t]
    \centering
    \caption[Bset]{Indirect ($\Gamma \Sigma$) and direct ($\Gamma \Gamma$) band gaps, $E_g$ (in eV), as well as effective masses $m$ 
(in units of the free electron mass $m_0$) calculated for holes and electrons in SL-Sb at relevant high-symmetry points of the Brillouin zone using the DFT (+SO) and TB (+SO) model presented in this work.
$m^1_{\Gamma}$ and $m^2_{\Gamma}$ stand for the effective masses of the light and heavy hole, whereas $m^x_{\Sigma}$ and $m^y_{\Sigma}$ denote anisotropic masses at $\Sigma$
calculated along the $\Gamma K$ and $\Gamma M$ directions, respectively.}
 \begin{tabular}{cccccccccccc}
      \hline
      \hline
   & &   &  &  \multicolumn{2}{c}{Holes} & & \multicolumn{4}{c}{Electrons} \\
\cline{5-6}
\cline{8-11}
 $\mathrm{Method}$ & $\ E^{\Gamma \Sigma}_g$ & $\ E^{\Gamma \Gamma}_g$ & & $\ m^1_{\Gamma}$ & $\ m^2_{\Gamma}$ &  & $\ m_{\Gamma}$ & $m^x_{\Sigma}$ & $\ m^y_{\Sigma}$ & $m_{K}$  \\
     \hline
  \    DFT        &  \   1.26          &  \  1.57       &  & \   0.08    & \  0.45  &  & \  0.09   &   \  0.14   &  \  0.45    & \ 0.39   \\
  \    TB         &  \   1.15          &  \  1.40       &  & \   0.06    & \  0.44  &  & \  0.06   &   \  0.13   &  \  0.42    & \ 0.36   \\
  \   DFT+SO      &  \   0.99          &  \  1.25       &  & \   0.10    & \  0.19  &  & \  0.08   &   \  0.14   &  \  0.46    & \ 0.40   \\
  \   TB+SO       &  \   0.92          &  \  1.14       &  & \   0.09    & \  0.11  &  & \  0.06   &   \  0.13   &  \  0.43    & \ 0.37   \\
      \hline                  
      \hline
    \end{tabular}
\label{masses}
    \end{table}

Let us now focus on the SO coupling in SL-Sb. Assuming a local character of the SO interaction, in the conventional atomiclike basis of $p$ orbitals, the SO Hamiltonian can be written 
as a sum of the intra-atomic contributions $H_{\mathrm{SO}}=\sum_jh^j$, each of which is given by \cite{Huertas}
\begin{equation}
h^j = \lambda \sum_{\sigma \sigma'} i (c^{\sigma \dag}_{z} \sigma_{\sigma\sigma'}^{x} c^{\sigma'}_{y} 
+ c^{\sigma \dag}_{z} \sigma_{\sigma \sigma'}^{y} c^{\sigma'}_{x}
+ c^{\sigma \dag}_{y} \sigma_{\sigma \sigma'}^{z} c^{\sigma'}_{x}
) + \mathrm{H.c.},
\label{Hso}
\end{equation}
where $\lambda$ is the intra-atomic SO coupling constant, $\sigma,\sigma'$ run over spin projections \{$\uparrow$, $\downarrow$\}, and $\sigma^{x}$, $\sigma^{y}$, $\sigma^{z}$ are the Pauli matrices.
After transformation of Eq.~(\ref{Hso}) to the basis of the WFs introduced above, the total TB+SO Hamiltonian of SL-Sb can be written as
\begin{equation}
H = H_{0} + \sum_{jmn}\sum_{pq} T^{(k_j)}_{mp} h^j_{pq} T^{(k_j)}_{nq},
\label{tot_hamilt}
\end{equation}
where $T^{(k_j)}$ is the sublattice-dependent matrix, determining the basis transformation, $|p^{(k_j)}_m\rangle = \sum_q T_{mq}^{(k_j)} |p_q \rangle$, explicitly given by Eq.~(\ref{Tmatrix}). $k_j$ is the sublattice index
of atom $j$, whereas $m$ and $n$ ($p$ and $q$) are the orbital indices running over 1,2,3 ($x,y,z$). 

\begin{figure}[tbp]
\includegraphics[width=0.47\textwidth, angle=0]{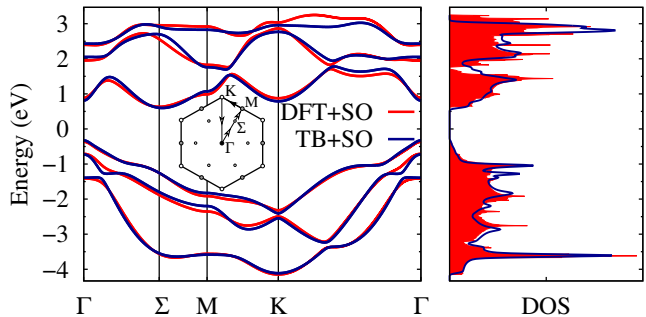}
\caption{Band structure (left) and density of states (right) calculated
including the SO coupling for SL-Sb using DFT and TB model given by Eqs.~(\ref{tb_hamilt}) and (\ref{Hso}).}
\label{SObands}
\end{figure}

In Fig.~\ref{SObands}, we show the relativistic electronic bands calculated from first-principles (DFT+SO) and using the TB+SO Hamiltonian [Eq.~(\ref{tot_hamilt})] obtained with $\lambda=0.34$ eV, which
is a fitting parameter quantitatively consistent with the intra-atomic SO strength of neutral Sb atoms \cite{Wittel}.
Both methods are in good agreement, which demonstrates the validity of the TB+SO Hamiltonian derived above.
The main effect of the SO coupling is the
band splitting in the vicinity of the crossing points. This effect is especially pronounced for the valence band at the $\Gamma$ point, and it results in a reduction of the indirect band gap
by $\sim$0.2--0.3 eV. From Table \ref{masses}, one can also see that SO significantly reduces the effective mass for the valence band. 
The position
and shape of the conduction band remains virtually unchanged.
Given that SL-Sb is a centrosymmetric crystal, each individual band remains doubly degenerate with respect to the spin degrees of freedom. The inversion symmetry, however, can be easily broken,
e.g., by an external electric field or by a substrate, which opens a way to induce spin splitting in SL-Sb and further reduce its energy gap.

We now turn to the problem of the Coulomb interaction in SL-Sb. 
In the static limit ($\omega=0$), the Coulomb interaction between lattice sites $i$ and $j$ can be defined 
in terms of the WFs $w_{i(j)}({\bf r})$ as
\begin{equation}
K_{ij} = \int d{\bf r} d{\bf r'} |w_{i}({\bf r})|^2 K({\bf r},{\bf r'}) |w_{j}({\bf r'})|^2,
\end{equation}
where $K({\bf r},{\bf r'})$ is the interaction, which in the absence of screening takes the form
$K=V({\bf r},{\bf r'})=e^2/|{\bf r}-{\bf r}'|$. The screening effects are taken into account at the level of the random phase approximation (RPA),
in which the reciprocal representation of the Coulomb interaction matrix is given by \cite{Aryasetiawan}
\begin{equation}
K({\bf q}) = \left[ 1 - V({\bf q})P({\bf q}) \right]^{-1}V({\bf q}),
\end{equation}
where $P({\bf q})$ is the static single-particle polarizability matrix, which in the WF basis reads \cite{Graf}
\begin{equation}
P_{mn}({\bf q})=\frac{1}{N_k}\sum_{{\bf k},ij}\frac{U^{{\bf k}*}_{mi}U^{{\bf k}'}_{mj}U_{ni}^{\bf k}U_{nj}^{{\bf k}'*}}{\varepsilon_i^{\bf k}-\varepsilon_j^{{\bf k}'}+i\eta},
\end{equation}
where ${\bf k}'={\bf k}+{\bf q}$, $U^{\bf k}_{mi}$ is a unitary transformation matrix defined in 
Eq.~(\ref{gauge}), $\varepsilon_i^{\bf k}$ is the eigenvalue of the full DFT Hamiltonian, $\eta$ is
a numerical smearing parameter, and the summation runs over the Brillouin zone involving transitions between
the occupied ($i$) and unoccupied ($j$) states only. For generality, we calculate $P({\bf q})$
including polarization
(i) by all relevant high-energy states, giving rise to the fully screened interaction 
$W(\bf q)$, and (ii) by the $p$-states only, which constitute the self-screened
interaction $U({\bf q})$.

\begin{figure}[t]
\includegraphics[width=0.47\textwidth, angle=0]{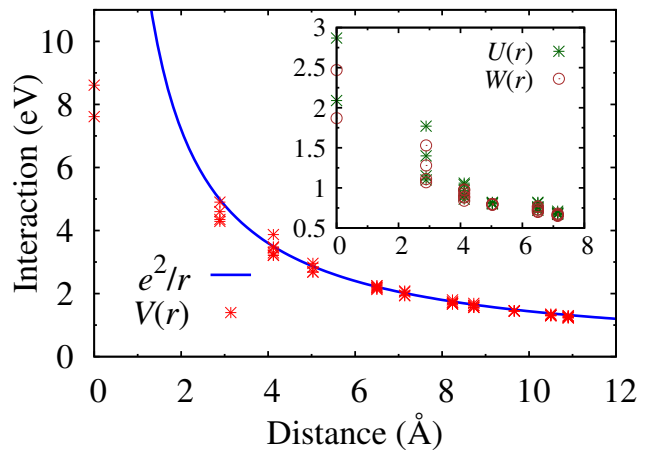}
\caption{Points: Bare Coulomb interaction $V(r)$ between the $p$ orbitals in SL-Sb calculated as a function of distance $r$ between the lattice sites.
Solid line: Classical Coulomb law $e^2/r$. Inset: Self-screened $U(r)$ and fully screened interactions $W(r)$ calculated within RPA.}
\label{U_fig}
\end{figure}

    \begin{table}[b]
    \centering
    \caption[Bset]{Bare ($V$), static self-screened ($U$), and static fully screened ($W$) Coulomb interactions (in eV) calculated between the $p$ orbitals in SL-Sb using RPA. 
Intersite interactions are averaged over the orbital indices ($m, n$) on each site and shown up to the fourth nearest neighbor (4NN).}
 \begin{tabular}{ccccccccc}
      \hline
      \hline
   &    \multicolumn{2}{c}{On-site ($i=j$)} & & \multicolumn{4}{c}{Intersite ($i\neq j$)} \\
\cline{2-3}
\cline{5-8}
                     &   $m=n$   &  $m\neq n$ & &     1NN    &     2NN    &     3NN    &   4NN       \\
     \hline
  \     $V$     \    & \  8.61 \ & \ 7.61   \ & & \  4.51  \ &  \ 3.41 \  &  \  2.78 \ &   \ 2.20  \ \\
  \     $U$     \    & \  2.87 \ & \ 2.09   \ & & \  1.32  \ &  \ 0.95 \  &  \  0.81 \ &   \ 0.77  \ \\
  \     $W$     \    & \  2.47 \ & \ 1.87   \ & & \  1.22  \ &  \ 0.91 \  &  \  0.80 \ &   \ 0.74  \ \\
      \hline                  
      \hline
    \end{tabular}
\label{U_tab}
    \end{table}

The calculated Coulomb interactions are shown in Fig.~\ref{U_fig} as a function of distance between the lattice sites, and also summarized in Table \ref{U_tab}. 
The bare interaction $V$ in SL-Sb is considerably smaller than that in graphene \cite{Wehling}, which is due to a more delocalized character of the 5$p$ orbitals
of Sb atoms, as well as due to the larger lattice constant of SL-Sb. Apart from the on-site bare interaction $V_{00}$, intersite interactions $V_{ij}$ are well
described by the classical Coulomb law $e^2/r$, as shown in Fig.~\ref{U_fig}. In contrast, the screening in SL-Sb is significantly stronger compared to graphene,
resulting in relatively weak interactions $U$ and $W$. Moreover, the screening originates predominantly from the polarization of 5$p$ orbitals,
whereas the contribution from high-energy states is not significant, thus making $U$$\sim$$W$.
In the context of the many-body lattice models (e.g., Hubbard model), the bare Coulomb interactions $V$ play, therefore, the role of an effective interaction entering 
the Hamiltonian. Mapping the nonlocal Coulomb interaction onto the local one \cite{Schuler}, it can be concluded that the leading terms of the kinetic and Coulomb energies in 
SL-Sb are comparable in magnitude, $|t_{01}|/(\bar{V}_{00}-\bar{V}_{01}) \sim 1.6$. This suggests a strongly correlated character of the 5$p$ electrons in SL-Sb.

To conclude, we have presented a systematic analysis of the electronic properties of single-layer antimony crystals. For this, we have performed relativistic first-principles calculations and derived an analytical TB model that describes the interactions between the 5$p$ single-particle states. 
We have shown that the strong spin-orbit coupling ($\lambda=0.34$ eV) plays an important role in the formation of the valence band, and can be used in conjunction with the electric field or substrate engineering to split the 
band degeneracy governed by the inversion symmetry. The TB model presented here accurately reproduces relativistic first-principles bands in a wide energy range and is flexible enough to describe a variety of experimental situations.
We have further calculated the strength of Coulomb interactions in this material and estimated the value of local (Hubbard) and intersite interactions, which is essential information to construct a many-body theory for this material. 
Importantly, the Coulomb screening is shown to be fully captured by the TB description, which makes the proposed model suitable for a comprehensive analysis of the electronic
properties, including large-scale simulations and many-body effects.
Our results also show indications of the strongly correlated character of electrons in SL-Sb, which can further stimulate theoretical and experimental interest in this 2D material.
\newline
\newline

%\begin{acknowledgments} 
The research has received funding from the European Union's Horizon 2020 Programme under Grant No.~696656 Graphene Core1, and from MINECO (Spain) through Grant No. FIS2014-58445-JIN.
%\end{acknowledgments} 

\end{document}